\documentstyle[amssymb,prl,aps,epsf]{revtex}

\newtheorem{thm}{Theorem}

\begin{document}
\draft
\title{Decoherence and Entanglement in Two-mode Squeezed Vacuum States}
\author{Tohya Hiroshima\cite{NEC1}}
\address{Fundamental Research Laboratories,\\
System Devices and Fundamental Research,\\
NEC Corporation, 34 Miyukigaoka, Tsukuba 305-8501, Japan}
\maketitle

\begin{abstract}
I investigate the decoherence of two-mode squeezed vacuum states by
analyzing the relative entropy of entanglement. I consider two sources of
decoherence: (i) the phase damping and (ii) the amplitude damping due to the
coupling to the thermal environment. In particular, I give the exact value
of the relative entropy of entanglement for the phase damping model. For the
amplitude damping model, I give an upper bound for the relative entropy of
entanglement, which turns out to be a good approximation for the
entanglement measure in usual experimental situations.
\end{abstract}


Quantum entanglement is an essential ingredient in quantum communication and
computation \cite{REV}. Therefore, it is of great importance to quantify the
entanglement to assess the efficacy of quantum information processing.
Recently, unconditional quantum teleportation of an unknown coherent state
has been realized experimentally by exploiting a two-mode squeezed vacuum
state as an entanglement resource \cite{FSBFKP} shortly after the
theoretical proposal of Ref.~\cite{BK1}. The two-mode squeezed vacuum state
shared between two parties - Alice and Bob - is formally generated from the
vacuum state $\left| vac\right\rangle $ by the unitary transformation, $%
U(r)=\exp \left[ -r\left( a_{1}^{\dagger }a_{2}^{\dagger }-a_{1}a_{2}\right)
\right] $, where $r(\geq 0)$ is called a squeezing parameter. The indices 1
and 2 refer to the optical mode of Alice and that of Bob, respectively. The
two-mode squeezed vacuum state $\left| \Psi \right\rangle =U(r)\left|
vac\right\rangle $ is written in the Fock basis as $\left| \Psi
\right\rangle =\left( \cosh r\right) ^{-1}\sum_{n=0}^{\infty }\tanh
^{n}r\left| n,n\right\rangle $, where $\left| n,n\right\rangle \equiv \left|
n\right\rangle _{1}\otimes \left| n\right\rangle _{2}$. Since $\left| \Psi
\right\rangle $ is a pure state, its amount of entanglement is uniquely
quantified by the von Neumann entropy of the reduced state of Alice or Bob:

\begin{equation}
S\left( \left| \Psi \right\rangle \right) =\cosh ^{2}r\log _{2}\left( \cosh
^{2}r\right) -\sinh ^{2}r\log _{2}\left( \sinh ^{2}r\right) .
\label{eq:ERPure}
\end{equation}
In real experimental situations, due to coupling to the environment, the
entangled state inevitably loses its purity; it becomes mixed. This
phenomenon - {\it decoherence} - is the most dangerous obstacle for all
entanglement manipulations. Several protocols for entanglement enhancement
or purification in continuous variable systems have been proposed \cite
{DGCZ1,OKW,PBP} and the decoherence of continuous variable states has been
studied by making use of Bell's inequality \cite{JLK}. However, the
quantification of entanglement of mixed entangled states in continuous
variable systems is still not well understood.

In this paper, I investigate the decoherence of two-mode squeezed vacuum
states by analyzing the relative entropy of entanglement \cite{VP}. I
consider two sources of decoherence separately; (i) {\it the phase damping}
and (ii) {\it the amplitude damping} due to the coupling to the thermal
environment. The damping is assumed to affect each mode of the state
independently with the same coupling parameters. In particular, I show that
the exact calculation of the relative entropy of entanglement can be
performed for the phase damping model. For the amplitude damping model, I
give an upper bound for the relative entropy of entanglement, which turns
out to be a good approximation for the entanglement measure in usual
experimental situations.

The relative entropy of entanglement of pure states reduces to the von
Neumann entropy of the reduced state of either subsystem. For a mixed state $%
\rho $ it is defined as $E_{R}(\rho )=\min_{\sigma \in {\cal D}}S(\rho
||\sigma )$, where $S(\rho ||\sigma )={\rm Tr}\left[ \rho \left( \log
_{2}\rho -\log _{2}\sigma \right) \right] $ is the quantum relative entropy.
The minimum is taken over ${\cal D}$, the set of all disentangled states. It
is usually difficult to calculate the relative entropy of entanglement for
mixed states, except for some specific states. Recently, the following
theorem on the relative entropy of entanglement has been proved \cite{WZ}
and it turns out to be quite suitable for my analysis.

\begin{thm}
\label{thm:WZ}

For a bipartite quantum state described by the density matrix of the form, 
\begin{equation}
\rho ={\sum_{n_{1},n_{2}}}a_{n_{1},n_{2}}\left| \phi _{n_{1}},\psi
_{n_{1}}\right\rangle \left\langle \phi _{n_{2}},\psi _{n_{2}}\right| ,
\label{eq:DMTheorem}
\end{equation}
the relative entropy of entanglement is given by 
\begin{equation}
E_{R}(\rho )=-{\sum_{n}\,}a_{n,n}\log _{2}a_{n,n}+{\rm Tr}\left( \rho \log
_{2}\rho \right) ,
\end{equation}
and the disentangled state $\rho ^{*}$ which minimizes the quantum relative
entropy $S\left( \rho ||\rho ^{*}\right) $ is 
\begin{equation}
\rho ^{*}={\sum_{n}\,}a_{n,n}\left| \phi _{n},\psi _{n}\right\rangle
\left\langle \phi _{n},\psi _{n}\right| .
\end{equation}
Here, $\left| \phi _{n}\right\rangle $ and $\left| \psi _{n}\right\rangle $
are orthonormal states of each subsystem.
\end{thm}

\begin{figure}
\begin{center}
\epsfxsize=8.0cm \epsfbox{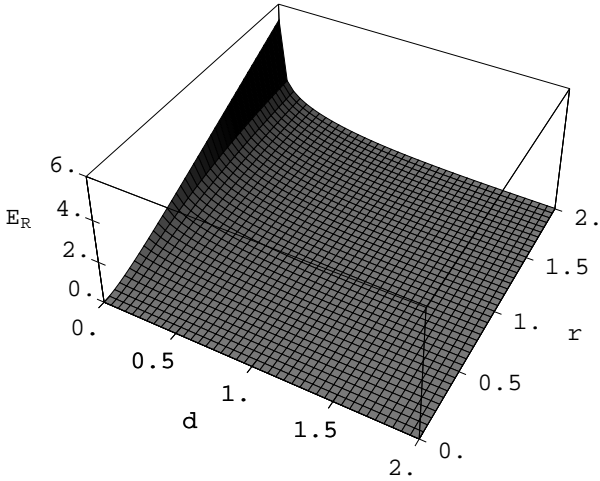}
%\epsfile{file=fig3.eps,width=8.0cm}
\end{center}
\caption{ The relative entropy of entanglement of the state of Eq.~(\ref
{eq:DMPhase}) as a function of squeezing parameter $r$ and the degree of
damping $d(=\gamma t)$. }
\label{fig:Phase}
\end{figure}

First, I consider the phase damping model. The density matrix obeys the
following master equation in the interaction picture, 
\begin{equation}
\frac{d}{dt}\rho (t)=\left( {\cal L}_{1}+{\cal L}_{2}\right) \rho (t),
\label{eq:Master}
\end{equation}
with 
\begin{equation}
{\cal L}_{i}\rho =\frac{\gamma }{2}\left[ 2a_{i}^{\dagger }a_{i}\rho
a_{i}^{\dagger }a_{i}-\left( a_{i}^{\dagger }a_{i}\right) ^{2}\rho -\rho
\left( a_{i}^{\dagger }a_{i}\right) ^{2}\right] .
\end{equation}
The solution of Eq.~(\ref{eq:Master}) with the initial condition $\rho
(t=0)=\left| \Psi \right\rangle \left\langle \Psi \right| $ is calculated as

\begin{equation}
\rho (t)=\frac{1}{\cosh ^{2}r}\sum_{n_{1},n_{2}=0}^{\infty }(\tanh
r)^{n_{1}+n_{2}}\exp \left( -\gamma t\left| n_{1}-n_{2}\right| ^{2}\right)
\left| n_{1},n_{1}\right\rangle \left\langle n_{2},n_{2}\right| .
\label{eq:DMPhase}
\end{equation}
It should be noted that the density matrix of Eq.~(\ref{eq:DMPhase}) takes
the form of Eq.~(\ref{eq:DMTheorem}) in Theorem \ref{thm:WZ}. Consequently,
it is possible to calculate exactly the relative entropy of entanglement of
the state of Eq.~(\ref{eq:DMPhase}). Figure~\ref{fig:Phase} shows the
relative entropy of entanglement $E_{R}$ thus computed as a function of
squeezing parameter $r$ and the degree of damping $d\equiv \gamma t$. In
numerical computations, the truncated photon number has been taken to be $%
\max \left( n_{1}\right) =\max \left( n_{2}\right) =100$, the value of which
is sufficiently large for numerical convergence. It is seen that with an
increasing amount of squeezing the amount of entanglement decreases rapidly
with the damping. For large values of $d$, $E_{R}$ is vanishingly small but
remains finite; it is still larger than the conceivable numerical errors. It
is not clear from the present numerical analysis whether the state is always
entangled for finite $\gamma t$.

\begin{figure}
\begin{center}
\epsfxsize=8.0cm \epsfbox{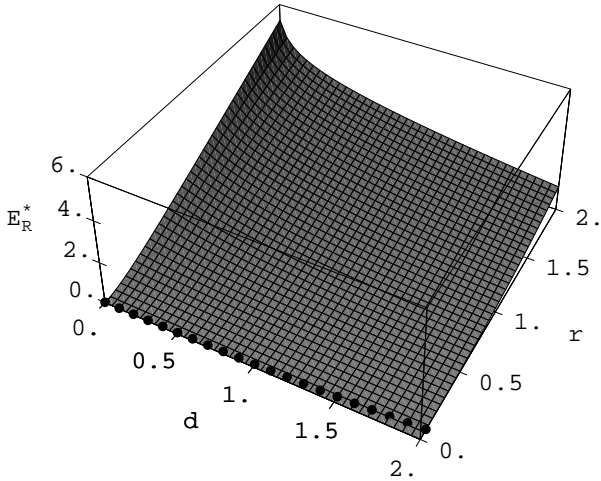}
%\epsfile{file=fig3.eps,width=8.0cm}
\end{center}
\caption{ The upper bound for the relative entropy of entanglement of the
state of Eq.~(\ref{eq:DMAmplitude}) as a function of squeezing parameter $r$
and the degree of damping $d(=\gamma t)$. The averaged number of photons in
the thermal environment $\overline{n}=0.01$. The dotted line is the
separability-inseparability border line. }
\label{fig:Amplitude1}
\end{figure}

\begin{figure}
\begin{center}
\epsfxsize=8.0cm \epsfbox{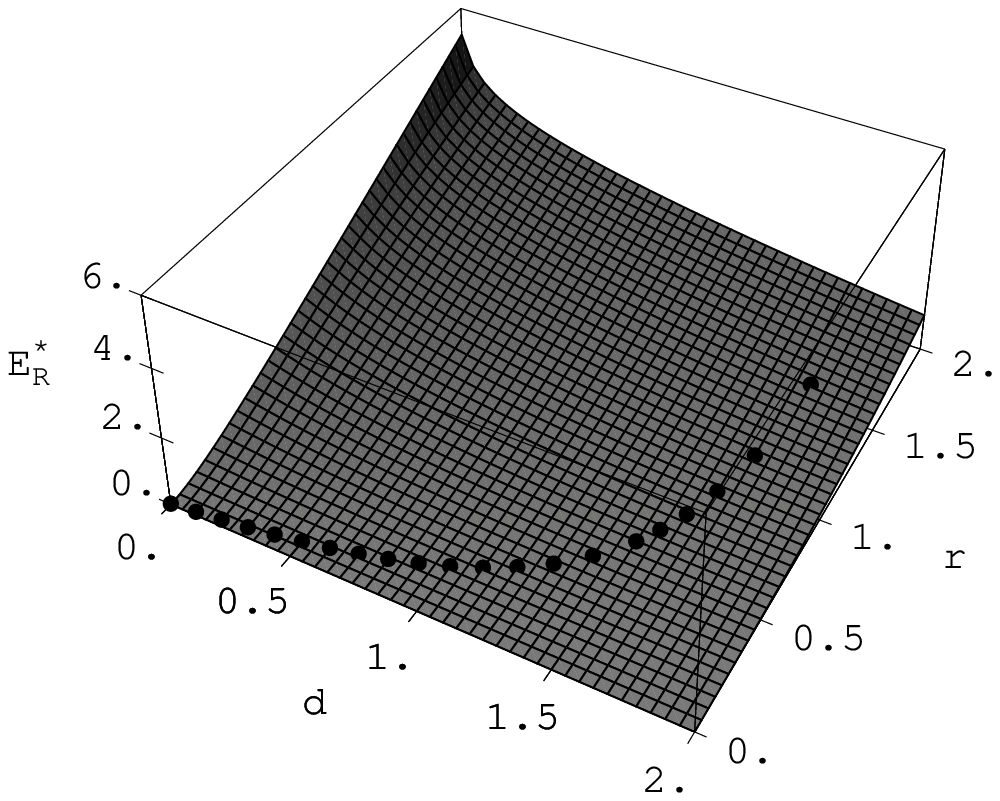}
%\epsfile{file=fig3.eps,width=8.0cm}
\end{center}
\caption{The same as Fig.~\ref{fig:Amplitude1}, but $\overline{n}=0.1$.}
\label{fig:Amplitude2}
\end{figure}

Next, I consider the amplitude damping model. The master equation for the
density matrix is the same as Eq.~(\ref{eq:Master}), but

\begin{equation}
{\cal L}_{i}\rho =\frac{\gamma }{2}\left( 1+\overline{n}\right) \left(
2a_{i}\rho a_{i}^{\dagger }-a_{i}^{\dagger }a_{i}\rho -\rho a_{i}^{\dagger
}a_{i}\right) +\frac{\gamma }{2}\overline{n}\left( 2a_{i}^{\dagger }\rho
a_{i}-a_{i}a_{i}^{\dagger }\rho -\rho a_{i}a_{i}^{\dagger }\right) ,
\end{equation}
where $\overline{n}$ is the average photon number of the thermal
environment. To solve the master equation, I first represent the two-mode
squeezed vacuum state as a continuous superposition of two-mode coherent
states \cite{JLK}: 
\begin{equation}
\left| \Psi \right\rangle =\int d^{2}\alpha G(\alpha ,r)\left| \alpha
,\alpha ^{*}\right\rangle ,
\end{equation}
where $G(\alpha ,r)=\exp \left[ -\left( \coth r-1\right) |\alpha
|^{2}\right] /\left( \pi \sinh r\right) $. According to the treatment of
Ref.~\cite{C}, the solution for the density matrix with the initial
condition $\rho (t=0)=\left| \Psi \right\rangle \left\langle \Psi \right| $
is calculated as 
\begin{equation}
\rho (t)=\int d^{2}\alpha \int d^{2}\beta G(\alpha ,r)G(\beta ,r)\sigma
_{1}\left( \alpha ,\beta \right) \sigma _{2}\left( \alpha ^{*},\beta
^{*}\right) ,
\end{equation}
where 
\begin{equation}
\sigma _{i}(\alpha ,\beta )=\frac{\left\langle \beta \right. |\left. \alpha
\right\rangle }{n(t)+1}\exp \left( -\xi _{1}\xi _{2}^{*}\right) D_{i}\left(
\strut {\xi }_{1}\right) \exp \left( \kappa a_{i}^{\dagger }a_{i}\right)
D_{i}^{\dagger }\left( \overline{\strut }\xi _{2}\right) .  \label{eq:Sigma}
\end{equation}
In Eq.~(\ref{eq:Sigma}), $n(t)=\overline{n}\left( 1-e^{-\gamma t}\right) $, $%
D_{i}\left( \xi _{j}\right) =\exp \left( \xi _{j}a_{i}^{\dagger }-\xi
_{j}^{*}a_{i}\right) $, and $\kappa =\ln \left\{ n(t)/\left[ n(t)+1\right]
\right\} $. Furthermore, ${\xi }_{1}$ and $\xi _{2}$ are given by ${\xi }%
_{1}=e^{-\frac{\gamma t}{2}}\left\{ \left[ n(t)+1\right] \alpha +n(t)\beta
\right\} /\left[ 2n(t)+1\right] $, and $\xi _{2}=e^{-\frac{\gamma t}{2}%
}\left\{ n(t)\alpha +\left[ n(t)+1\right] \beta \right\} /\left[
2n(t)+1\right] $. By noting the completeness of coherent states, it is
straightforward to show that

\begin{eqnarray}
&&_{i}\left\langle m_{1}\right| D_{i}\left( \strut {\xi }_{1}\right) \exp
\left( \kappa a_{i}^{\dagger }a_{i}\right) D_{i}^{\dagger }\left( \xi
_{2}\right) \left| m_{2}\right\rangle _{i}=\frac{1}{\pi ^{2}}\sqrt{\frac{1}{%
m_{1}!m_{2}!}}\int d^{2}\gamma _{1}\int d^{2}\gamma _{2}\gamma
_{1}^{m_{1}}\left( \gamma _{2}^{*}\right) ^{m_{2}}  \nonumber \\
&&\times \exp \left[ -\frac{1}{2}\left( |\gamma _{1}|^{2}+|\gamma
_{2}|^{2}+|\gamma _{1}-{\xi }_{1}|^{2}+|\gamma _{2}-\xi _{2}|^{2}\right)
+\left( \gamma _{1}-{\xi }_{1}\right) ^{*}\left( \gamma _{2}-\xi _{2}\right)
e^{\kappa }+\sqrt{-1}%
\mathop{\rm Im}%
\left( {\xi }_{1}\gamma _{1}^{*}\right) -\sqrt{-1}%
\mathop{\rm Im}%
\left( \xi _{2}\gamma _{2}^{*}\right) \right] .  \label{eq:Elem1}
\end{eqnarray}
Using Eq.~(\ref{eq:Elem1}), we obtain 
\begin{eqnarray}
&&\left\langle n_{1},n_{3}\right| \rho (t)\left| n_{2},n_{4}\right\rangle =%
\frac{R}{\pi ^{4}\sinh ^{4}r}\sqrt{\frac{1}{n_{1}!n_{2}!n_{3}!n_{4}!}}\int
d^{2}\gamma _{1}\int d^{2}\gamma _{2}\int d^{2}\gamma _{3}\int d^{2}\gamma
_{4}\gamma _{1}^{n_{1}}\left( \gamma _{2}^{*}\right) ^{n_{2}}\gamma
_{3}^{n3}\left( \gamma _{4}^{*}\right) ^{n_{4}}  \nonumber \\
&&\times \exp \left[ -|\gamma _{1}|^{2}-|\gamma _{2}|^{2}-|\gamma
_{3}|^{2}-|\gamma _{4}|^{2}\right] \exp \left[ P\left( \gamma _{1}^{*}\gamma
_{3}^{*}+\gamma _{2}\gamma _{4}\right) +Q\left( \gamma _{1}^{*}\gamma
_{2}+\gamma _{3}^{*}\gamma _{4}\right) \right] ,  \label{eq:Elem2}
\end{eqnarray}
where 
\begin{equation}
P=R\,e^{-\gamma t}\coth r,
\end{equation}
\begin{equation}
Q=\frac{1}{n(t)+1}\left\{ n(t)+R\,e^{-\gamma t}\left[ n(t)+1-e^{-\gamma
t}\right] \right\} ,
\end{equation}
and 
\begin{equation}
R=\left\{ \coth ^{2}r\left[ n(t)+1\right] ^{2}-\left[ n(t)+1-e^{-\gamma
t}\right] ^{2}\right\} ^{-1}.
\end{equation}
The integral of Eq.~(\ref{eq:Elem2}) can be evaluated by expanding $\exp
\left[ P\left( \gamma _{1}^{*}\gamma _{3}^{*}+\gamma _{2}\gamma _{4}\right)
+Q\left( \gamma _{1}^{*}\gamma _{2}+\gamma _{3}^{*}\gamma _{4}\right)
\right] $ with respect to $\gamma _{1}^{*}\gamma _{3}^{*}$, $\gamma
_{2}\gamma _{4}$, $\gamma _{1}^{*}\gamma _{2}$, and $\gamma _{3}^{*}\gamma
_{4}$. Finally we obtain 
\begin{eqnarray}
\rho (t) &=&\sum_{n_{1},n_{2}=0}^{\infty }c_{n_{1},n_{2}}^{(0)}\left|
n_{1},n_{1}\right\rangle \left\langle n_{2},n_{2}\right|  \nonumber \\
&&+\sum_{k=1}^{\infty }\sum_{n_{1},n_{2}=0}^{\infty
}c_{n_{1},n_{2}}^{(k)}\left| n_{1},n_{1}+k\right\rangle \left\langle
n_{2},n_{2}+k\right| +\sum_{k=1}^{\infty }\sum_{n_{1},n_{2}=0}^{\infty
}c_{n_{1},n_{2}}^{(k)}\left| n_{1}+k,n_{1}\right\rangle \left\langle
n_{2}+k,n_{2}\right| ,  \label{eq:DMAmplitude}
\end{eqnarray}
where 
\begin{equation}
c_{n_{1},n_{2}}^{(k)}=\frac{\sqrt{n_{1}!n_{2}!(n_{1}+k)!(n_{2}+k)!}}{\sinh
^{2}r}P^{n_{1}+n_{2}}Q^{k}R\sum_{l=0}^{\min (n_{1},n_{2})}\frac{1}{%
l!(l+k)!(n_{1}-l)!(n_{2}-l)!}\left( \frac{Q}{P}\right) ^{2l}.
\end{equation}
To estimate the relative entropy of entanglement of the state $\rho (t)$ of
Eq.~(\ref{eq:DMAmplitude}), I write $\rho (t)$ as a convex combination of
new density matrices, $\rho _{0}(t)$ and $\rho _{k}^{(\pm )}(t)$ $%
(k=1,2,...) $:

\begin{equation}
\rho (t)=p_{0}\rho _{0}(t)+\sum_{k=1}^{\infty }p_{k}^{(+)}\rho
_{k}^{(+)}(t)+\sum_{k=1}^{\infty }p_{k}^{(-)}\rho _{k}^{(-)}(t).
\label{eq:Convex}
\end{equation}
The first, second, and third terms of the right-hand side of Eq.~(\ref
{eq:Convex}) correspond to the first, second, and third terms of the
right-hand side of Eq.~(\ref{eq:DMAmplitude}), respectively. By convexity of
the relative entropy of entanglement \cite{VP}, we have that 
\begin{equation}
E_{R}\left( \rho (t)\right) \leq p_{0}E_{R}\left( \rho _{0}(t)\right)
+\sum_{k=1}^{\infty }p_{k}^{(+)}E_{R}\left( \rho _{k}^{(+)}(t)\right)
+\sum_{k=1}^{\infty }p_{k}^{(-)}E_{R}\left( \rho _{k}^{(-)}(t)\right) .
\label{eq:ERMixed}
\end{equation}
The density matrices $\rho _{0}(t)$ and $\rho _{k}^{(\pm )}(t)$ $(k=1,2,...)$
have the form of Eq.~(\ref{eq:DMTheorem}) in Theorem \ref{thm:WZ}, so the
right-hand side of Eq.~(\ref{eq:ERMixed}) can be calculated exactly and it
yields an upper bound for the relative entropy of entanglement of $\rho (t)$%
. At $\gamma t=0$, this upper bound $E_{R}^{*}$ gives the exact value of the
relative entropy of entanglement of Eq.~(\ref{eq:ERPure}). Figures \ref
{fig:Amplitude1} and \ref{fig:Amplitude2} show $E_{R}^{*}$ thus computed as
a function of squeezing parameter $r$ and the degree of damping $d(\equiv
\gamma t)$. In Fig.~\ref{fig:Amplitude1} (\ref{fig:Amplitude2}), $\overline{n%
}$, the averaged number of photons in the thermal environment, has been
taken to be 0.01 (0.1). The dotted line on the data surface in each figure
indicates the separability-inseparability border line, which is given by the
necessary and sufficient separability criterion for the two-mode squeezed
state in the thermal environment \cite{DGCZ2,S}; $\gamma t=\ln \left[
1+\left( 1-e^{-2r}\right) /\left( 2\overline{n}\right) \right] $. Within the
inside region (above the dotted line) the state is entangled, while within
the outside region (below the dotted line) it is separable and the relative
entropy of entanglement $E_{R}$ vanishes. It is seen that the values of the
upper bound $E_{R}^{*}$ on these border lines for $r\lesssim 1$ are already
negligibly small. Although $E_{R}^{*}$ is merely an upper bound for $E_{R}$,
it may be considered as an approximation for $E_{R}$. Since $r\lesssim 1$
and $\overline{n}\ll 1$ in usual experimental situations, this approximation
is a fairly good one in the sense that the upper bound $E_{R}^{*}%
\thickapprox 0$ for the separable region.

In summary, the decoherence of two-mode squeezed vacuum states has been
investigated by analyzing the relative entropy of entanglement for (i) the
phase damping model and (ii) the amplitude damping model. For the phase
damping model, a method for the exact numerical computation of the relative
entropy of entanglement has been established. For the amplitude damping
model, a good approximation for the relative entropy of entanglement in
usual experimental situations has been introduced.

\end{document}